\documentclass[AMA,LATO1COL]{WileyNJD-v2}

\articletype{Article Type}%

\received{26 April 2016}
\revised{2 June 2016}
\accepted{6 June 2016}

\begin{document}

\title{Reliable Access to Massive Restricted Texts: Experience-based Evaluation}

\author[1]{Zong Peng*}

\author[1]{Beth Plale}

%\author[3]{Author Three}

\authormark{AUTHOR ONE \textsc{et al}}

\address[1]{\orgdiv{School of Informatics, Computing, and Engineering}, \orgname{Indiana University Bloomington}, \orgaddress{\state{Indiana}, \country{US}}}

%\address[2]{\orgdiv{Data To Insight Center}, %\orgname{Indiana University}, \orgaddress{\state{Indiana}, %\country{US}}}

%\address[3]{\orgdiv{Org Division}, \orgname{Org Name}, \orgaddress{\state{State name}, \country{Country name}}}

%\corres{*Corresponding author name, This is sample corresponding address. \email{authorone@gmail.com}}

%\presentaddress{This is sample for present address text this is sample for present address text}

\abstract[Summary]{Libraries are seeing growing numbers of digitized textual corpora that frequently come with restrictions on their content.  Computational analysis corpora that are large, while of interest to scholars, can be cumbersome because of the combination of size, granularity of access, and access restrictions. Efficient management of such a collection for general access especially under failures depends on the primary storage system. In this paper, we identify the requirements of managing for computational analysis a massive text corpus and use it as basis to evaluate candidate storage solutions.  The study based on the 5.9 billion page collection of the HathiTrust digital library.  Our findings led to the choice of Cassandra 3.x for the primary back end store, which is currently in deployment in the HathiTrust Research Center.}


\jnlcitation{\cname{%
\author{Williams K.}, 
\author{B. Hoskins}, 
\author{R. Lee}, 
\author{G. Masato}, and 
\author{T. Woollings}} (\cyear{2016}), 
\ctitle{A regime analysis of Atlantic winter jet variability applied to evaluate HadGEM3-GC2}, \cjournal{Q.J.R. Meteorol. Soc.}, \cvol{2017;00:1--6}.}

\maketitle

%\footnotetext{\textbf{Abbreviations:} ANA, anti-nuclear antibodies; APC, antigen-presenting cells; IRF, interferon regulatory factor}

\section{Introduction}\label{sec1}

Libraries are increasingly confronted with digitized textual corpora that can be large. For instance, the Mass Digitization \cite{coyle2006mass} project is attempting to digitize every book ever printed. The New York Public Library has a living collection of 757,072 digitized materials as of February 2019~\footnote{https://digitalcollections.nypl.org/}.  Provisioning computational analysis (e.g., text mining, natural language processing) of a large digitized collection of textual content poses a number of challenges.  First is the size of the collection. The HathiTrust digital library \cite{york2010building} 
holds 5.9 billion individually digitized pages, and page level granularity in access has been stated by researchers of the HathiTrust collection as a need. Provisioning computational analysis is further complicated by access restrictions on the content based on donor wishes, copyright, etc.  A digitized textual corpus will contain a mix of public domain and in-copyright material if the content is drawn from library sources and is bigger than a homogeneous corpora (e.g., collection of letters of author Ray Bradbury). 

Computational analysis of texts that respects the legal restrictions on the texts has come to be called \textit{non-consumptive research}. Acceptable research, under non-consumptive research, is restricted to computational analysis that is performed on one or more volumes (digitized texts); unacceptable research, on the other hand, is research in which a human being reads or displays substantial portions of an in-copyright or rights-restricted volume to understand expressive content presented within the volume.  The term "non-consumptive research" originates in the Authors Guild, Inc. and Association of American Publishers, Inc. \textit{et al.} vs Google Inc. Amended Settlement Agreement~\footnote{Amended Google Settlement, at the Internet Archive Wayback Machine (web.archive.org) by date of Dec 8, 2011 and URL of http://www.googlebooksettlement.com/Amended-Settlement-Agreement.zip} filed with the U.S. District Court Southern District of New York in 2009. The Amended Settlement Agreement ultimately failed, but the notion of non-consumptive research was continued.  The term "non-consumptive research" is defined in section 1.93 of the Settlement Agreement. 

Serving a large digitized textual corpus for analytical analysis is far more than managing the digitized texts themselves.  Related content must also be managed including rights information, bibliographic (semantic), and structural metadata.  The heterogeneous content needs to seamlessly interoperate for efficient data access while adhering to the policy of non-consumptive research. Addtionally, the types of data involved have different access patterns. Whereas semantic metadata is queried, reads of textual content are frequently random access for reasons we give later. This heterogeneity of needs suggests a polyglot architecture and persistence~\cite{Sharp:2013:DAH:2566943}. Our study is focused on the optimal storage solution for a large digitized textual collection in the context of a polyglot architecture.   

In choosing an optimal storage solution (which we'll also refer to as a "database") for large digitized textual corpora, it is frequently the performance of the database that is the major decisive factor.  This has been the conclusion of a number of studies including Louren\c{c}o \textit{et al.} \cite{lourencco2015choosing} who synthesizes conclusions from earlier comparative evaluations of SQL and noSQL data storage systems including ~\cite{li2013performance}.  Konstantinou \textit{et al.} \cite{konstantinou2011elasticity} conclude that HBase provides the highest throughput compared to Cassandra and Riak while Rabl \textit{et al.} \cite{Rabl:nosql} found Cassandra's throughput to be consistently better than Voldemort, HBase, Redis, VoltDB and MySQL.  Louren\c{c}o \textit{et al.} \cite{lourencco2015choosing} and Kashyap \textit{et al.} \cite{kashyap2013benchmarking} both note that NoSQL is constantly evolving and most evaluations quickly become obsolete. Rabl \textit{et al.} \cite{Rabl:nosql} compares NoSQL and SQL datastores using the YCSB synthetic workload. While instructive, the access pattern used in the workloads and the randomly generated synthetic data are not representative of digitized books. 
 
  In addition to performance, however, there are qualitative features that matter in making a choice: the data model, distribution architecture, and failure resistance. 
  
The three databases that we study are all NoSQL databases.  While relational databases are strong and reliable solutions for storing massive textual data, they were not included in our study because of the unique demands of our application:   5.9 billion pages, a largely single-writer, multiple reader access model, and wide variability in page counts across the books of the collection. The size of the collection (6.1 billion pages) demands a multi-node solution and RDBMS' scalability is limited by the strong ACID properties \cite{han2011survey, rabl2012solving} which are an infrequent need in our single-writer, multiple reader setting.  Too, books vary widely on their page counts. While pages can be broken out in a table, the table will be subject to significant join activity which will reduce performance. A common query type is the full text query, which, like noSQL, cannot be efficiently supported by RDBMS and requires a separate index, making many of the indexing benefits of RDBMS unavailable. 

We additionally do not consider a filesystem solution for a similar reason:  6.2 billion files stresses the availability, scaling and resilience features of commodity nodes.

This paper is a evaluation of three database solutions for managing a massive digital book collection\footnote{We use the terms "book", "item" and "volume" interchangeably to refer to a coherent set of pages with a beginning and an end}. We illustrate the environment through a model, define the requirements based on the HathiTrust Research Center~\footnote{www.hathitrust.org/htrc}, and use the requirements to determine an optimal data storage system. 

The remainder of this paper is organized as follows. Section \ref{sec:problemSpace} discusses the challenges of managing restricted digitized textual content. Section \ref{sec:architecture} proposes an architectural model to address the challenges. Section \ref{sec:datastore} analyzes the data characteristics and access pattern for digitized volumes and their metadata and narrows down the scope of different types of data stores that can match the data/metadata. Sections \ref{sec:qualitative} and  \ref{sec:quantitative} are comparative analyses of possible choices. Then section \ref{discussion} summarizes the comparative analysis. Section \ref{sec:usecase} draws on a real use case from HathiTrust~\footnote{HathiTrust http://hathitrust.org} and describes the architectural model's fit in that environment. We conclude with future work.

\section{Motivation}\label{sec:problemSpace}

The recent availability of historical digitized textual corpora is a strong motivator for researchers who can manipulate text mining tools to discover trends \cite{michel2011quantitative} and run statistical  models\cite{bamman2014bayesian} on texts that were not previously amenable to computational analysis.   Provisioning access to a collection of millions of digitized books, many of which are under copyright, requires that researchers interact through remotely hosted tools and queries as they cannot download the content to their desktops.  By Foster ~\cite{ResearchInfraIanFoster}, the infrastructure for interaction with a large digitized textual corpora is a \emph{remote secure enclave}, where physical  constraints  are placed on  data  access  and  export.  Foster differentiates between air-gapped enclaves and secure remote access enclaves. In the latter, "the analyst connects remotely, for example over a virtual private network, to the secure enclave. 

The types of information that must be managed by the HathiTrust Research Center in provisioning access to the digital texts is as follows:

\begin{itemize}
\item \textit{Digitized texts}.  Digitized texts constitute the significant majority of content in terms of bytes.  The books in the HathiTrust vary in size from a few tens of pages to thousands of pages, with an average volume length of 350 pages. With a corpus of nearly 16 million volumes, billions of pages must be managed. As shown earlier, the dominant read access pattern is random, and access within a volume is always random. Writing is much less frequent and when it occurs, does so as full-volume replace. Read access can be at the volume or page level. 

\item \textit{Rights information.} Rights information is small in size, though the dominant access pattern is also random read based on the identifier of the digitized volume. Rights information is updated frequently. 

\item \textit{Semantic metadata}. Complex metadata, coded by librarians using library-specific metadata formats (i.e., METS, MARC); used to describe digitized unstructured content. Semantic metadata is drawn primarily from the bibliographic record of a book as determined by the contributing library, though can be augmented with additional metadata from other sources. 
Semantic metadata includes mostly bibliographic information such as author names, title, etc.  The primary content that is used in managing the digital corpus is in the MARC\footnote{https://www.loc.gov/marc/} record form.

\item \textit{Structural metadata}. Structural metadata describes the structure of the digital object: number of pages, size of each page, page checksum, etc. It also comes from the contributing library in the form of a METS record.  This information is used to validate the integrity of the digitized content. 
\end{itemize}

The heterogeneous data asset has several additional characteristics worth noting:  First, \textit{copyright is a legal term, not a management strategy}.  The copyright status of a text is bound by jurisdiction. One book may be bound by US copyright law, and another by European copyright law.  A researcher in country $\mathbf{X}$ may have access rights that differ from researcher in country $\mathbf{Y}$. In addition, the copyright status of a book may change over time due to death, copyright period, orphan status where copyright status cannot be determined, further complicating management.

Second, \textit{collection layout is not optimized for computational analysis}. The HathiTrust digital library resides in a filesystem and is organized according to contributing organization (research library).  It uses the deep directory hierarchy organization called a Pairtree\footnote{https://confluence.ucop.edu/display/Curation/PairTree}. This pairtree is duplicated to another location (our organization) via periodic rsync.  The data are thus organized as a tree with the main branches allocated to the contributing libraries.  This organization facilitates addition of content, because additions are by contributing library. But this contributing-library organization fails to reflect any known locality patterns by researchers interested in computational analysis (e.g., genre, publication date, author).  As a result, all read requests are effectively random access. 

Finally, the \textit{workload exhibits bifurcation}.  A read request from a researcher is a request for a batch of items described by a workset of volumes a researcher is studying.  This workset is observed to be either 1,000 or so books for the researcher who is drawing from the works they have personally read, or in the range of 300,000 books for the computationally astute researcher who may, for instance, want to mine all the digitized content with Library of Congress subclass of Q (science texts.)  Both ends of this bifurcated workload must be handled by the storage system.  Further, client requests can be either volume ID based or feature based. For feature-based requests, semantic metadata sources need to be consulted. In all cases, backend complexity should be transparent to users. 

The process to request digitized volumes usually touches all types of content through a multi-step process: 1) query semantic metadata to retrieve volume identifiers; 2) validate rights information and 3) retrieve textual content and structural metadata. 

\subsection{System Architecture} \label{sec:architecture}
\begin{figure}
\centering
\includegraphics[ scale=1.2]{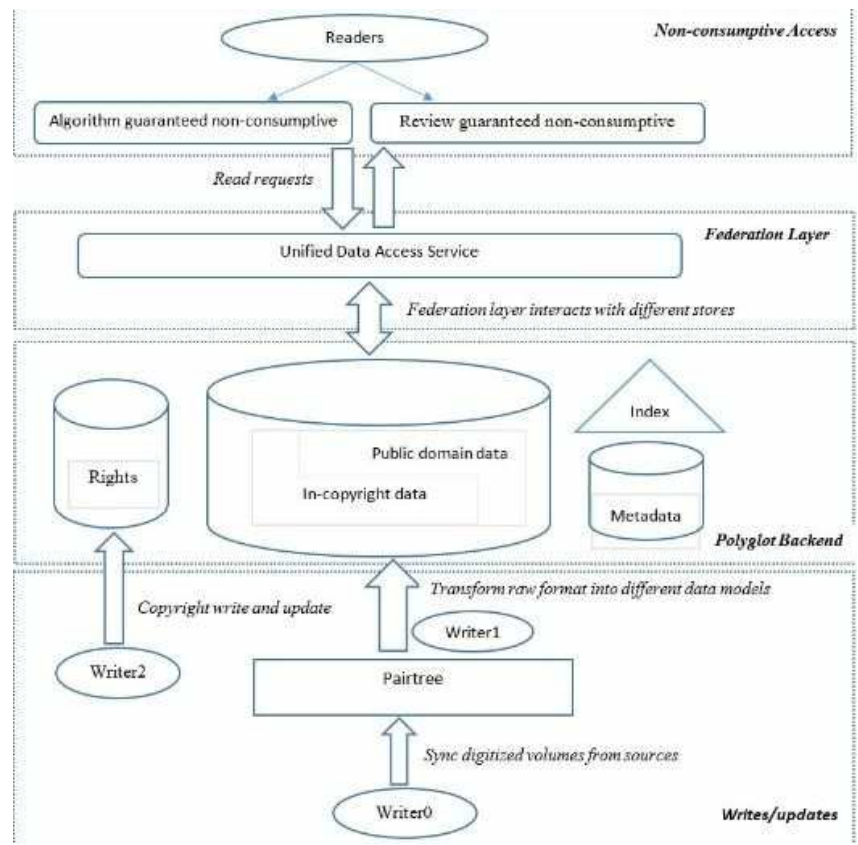} 
%\vspace*{-0.8cm}
\caption{Architectural model of HathiTrust Research Center. Writer0 synchronizes digitized volumes on disk along with metadata from sources; writer1 writes raw data/metadata into different stores; Writer2 updates copyright information periodically. Between the databases and the two types of readers is a complexity-hiding layer.}
\label{fig:htrcArch}
\end{figure}

A model for data management is captured in Figure \ref{fig:htrcArch} and described below.  The model can be read bottom to top, with ingest source at the bottom of the figure as $Writer0$ and $Writer1$ and researcher's computational interaction at the top as $Readers$.

\textbf{Ingest}
Page level data reside on disk in a Pairtree organization with a deep file system hierarchy having a single root. The next level down from the root are major branches that HathiTrust organizes into per library branches. That is, each major branch holds digitized content that has been contributed by one academic or research library. There is one major branch per contributing library. 

Raw digitized books/pages in Pairtree form and accompanying semantic metadata are streamed from separate writers and updated periodically when new OCR pages for instance are produced with better quality or for newly digitized books. Copyright information is ingested and updated periodically from a separate authoritative source. Update of rights information is done frequently in response to active monitoring by the provider of copyright status changes. 

\textbf{Back End Storage}
 This Pairtree organization and the page-level file granularity (6.2 billion files) makes disk storage less than ideal. The small files can overwhelm a file system; the way in which the files are organized on disk does not have bearing on the content-based retrievals that text mining researchers will have.   In other words, contributing library is meaningless to researchers who are analyzing texts. 
 
The large text corpus might be split into the public domain portion and the in-copyright portion.   But copyright status depends not on the the country in which the text is location, but on the country in which the user is located. For example, a public domain volume with rights information ''$PDUS$'' is in the public domain in US, which means for users outside the US, the volume is not in the public domain. Too, texts will have their status changed; this is more frequent an occurrence with the HathiTrust collection than one might imagine. 

Rights information has an access pattern that is different from the main corpus, thus rights are stored separately as shown in Figure \ref{fig:htrcArch}. For an incoming user request to a batch of volumes, first the rights information is checked against user access level.  Rights information is of small size and needs to be checked frequently so requires low latency. 

Semantic metadata is typically searched on multiple bibliographic fields. Search against semantic metadata is a primary way in which users find volume identifiers. Most of queries against semantic metadata are to the indexed full text. Structural metadata can be used to validate textual content for user requests and detect bit rot on disk. 

\textbf{Federation Layer}
Since different types of data can be managed by multiple different storage engines used in backend, each of which designed for a specialized data and access pattern, interaction with multiple storage engines efficiently by providing users with a unified data access service to hide backend complexity is preferred. Users only need to interact with the standard interface when implementing their custom analysis algorithms. Any future changes of data management systems in the backend would not require re-implementation of user algorithms.

\textbf{Non-consumptive Access}
Non-consumptive use of digitized books refers to computational analysis of the copyrighted content that is carried out in such a way that human consumption of texts is prohibited. One way to implement non-consumptive access is to use off-the-shelf analysis tools that have been certified to not return copyrighted content.  Another is to provide a researcher with a secure environment where she can install her tools and run them.  The Data Capsule service ~\cite{zeng2014cloud} is used in HTRC to provide non-consumptive access to the HathiTrust collection.   A user checks out a Capsule for a period of time.   Network ports are opened to allow installation, and then closed when the analysis tools actually access the data.  All results leaving the Capsule are subject to manual review.  

\section{Storage Alternatives}\label{sec:datastore}

Because of the sheer size of the textual corpus, the differing rates of update, and access patterns, there is no one-size-fits-all storage system that can manage the content, which suggests a polyglot backend. We thus undertake to study the different types of data stores to select the best fit in this model. 

\subsection{Primary Data Storage}
Big data applications have revolutionized the underlying infrastructure for the storage system. High availability, good scalability, and fast key value access have become standard requirements for storage systems.  For reasons given earlier, we chose to compare across several NoSQL options.    A NoSQL store is a distributed storage system that organizes data in a schema-less format and relaxes the consistency constraint in RDBMS in order to provide high availability, automatic scaling, resilience under failure, etc. NoSQL stores are generally classified into four groups: key-value store, document store, column-oriented store and graph databases \cite{hecht2011nosql}. The document partitioned inverted index is considered a special kind of document store. Digitized items are standalone entities except when they participate in a collection (such as issues in a journal).  Thus we eliminate from our study the graph store with its model of nodes and connecting edges, and focus on the key-value store, document store, and column-oriented store.

The options considered are as follows:

\textbf{Key-value store.}
The Key-value store has a comparatively simple data model. Rights information of digitized books maps each identifier of a volume to the access level information (in short string), which fits key value model. It is updated frequently and is accessed for all requests before reading textual content. Low latency is important so that rights validation is not the performance bottleneck. Because of the small size of rights information, memory based solution in addition to key-value store is ideal. For example, Redis \cite{carlson2013redis}, an in-memory but persistent on disk database, can be good option. However, key-value model is too restricted for digitize volumes with rich metadata.

\textbf{Document store.}
Each document in document store has a set of key value pairs with no fixed schema. Document stores are hierarchical in nature because documents can be embedded. Document model actually fits digitized books and structural metadata very well if each digitized volume is seen as a document with <pageSequence, pageContent> as key value pairs. Structural metadata for each digitized volume can also be stored as extra key value pairs in the same document. The most popular representative is MongoDB. 

\textbf{Inverted index datastore.}
Inverted index is the most efficient state of the art data structure for full text search. Data is also organized in document-like conceptual view with all fields indexed. It is ideal to both store and index semantic fields with inverted index. 

\textbf{Column-oriented store.}
The data model of column-oriented stores fit digitized volumes and their structural metadata as well. It gives a conceptual view of a big table with flexible schema where OCR pages and structural metadata fields can be seen as columns. Similar with document store, column-oriented stores are also widely used for big data applications. The popular representatives are HBase and Cassandra. 

\subsection{Federated Access Layer}
Multiple storage systems are needed for heterogeneous types of data, suggesting a polyglot solution
for seamless access to heterogeneous storage solutions. While there are benefits of a polyglot architecture, its problems are also well known \cite{sellami2014odbapi}. To make the back end complexity transparent to users, provide an unified user interface, and enforce best practice for optimal query efficiency, a federated layer between polyglot backend and non-consumptive clients are needed. This federated layer needs to 1) expose a unified interface that is backend agnostic; 2)optimize the client queries into an optimal plan to enforce the best practice based on characteristics
of different data stores; 3) and be distributed, highly available and efficient so that no single point of failure is introduced and the cost of an extra layer is minimized.   The three needs of the federated layer requires mapping several data models into a single unified model, generating an optimal query plan for the multiple NoSQL stores, and making the federated access layer itself distributed and failure resistant. 

Different data models can be either aggregated into a unified self-defined resource model such as ODBAPI resource model \cite{sellami2014odbapi} or mapped to relational model such as Apache Calcite\footnote{https://calcite.apache.org} that further enables pluggable custom query plan to optimize query performance against polyglot backend. In the case of HathiTrust, queries that read the textual content of volumes with filtering criteria on bibliographic metadata and rights information requires the execution of metadata-based sub-queries to find the list of identifiers prior to reading primary storage system to minimize the amount of data read and avoid big joins. Queries are typically sent through REST methods (e.g. GET, PUT, etc) and unified access service should be replicated for both functionality redundancy and workload distribution. 

\section{Database Feature Comparison}\label{sec:qualitative}

In HathiTrust Research Center, a single large collection dwarfs all other supporting databases. Thus the data storage system of the large collection will determine the overall performance of entire system. Both semantic metadata and rights information are comparatively small and can be served by memory based solutions with low latency.  The overall system performance of most queries largely relies on the primary data store for digitized volumes. 

Both \textit{document store} and \textit{column-oriented store} can accommodate the data model and access pattern requirements for digitized volumes and structural metadata. For document store, \textit{MongoDB} is chosen as representative. For column-oriented store, we choose \textit{HBase} and \textit{Cassandra} because Cassandra is strictly not a pure column-oriented store but a hybrid of key-value and column oriented store. To find the optimal solution, more detailed requirements with comparative study (both qualitative and quantitative) for main storage systems are needed. 

We identified six requirements for a data storage system that manages the Big Data textual data of HathiTrust with its access restrictions.   They are as follows:   

\textbf{High read throughput for random access}
As explained earlier, for our use case, the workload against the primary data store can be considered random access.  Additionally,  client requests follow a bifurcated workload pattern: a typical request will be for about 1,000 volumes for the user that is  analyzing texts that are part of the corpus of books they have actually read; or for the digital humanities researcher that is branching out into Big Data, the request will have a size of about 300,000 volumes.  This is, for instance, the size of all volumes in HathiTrust in the Library of Congress Q subclass (science texts, see https://www.hathitrust.org/visualizations\_callnumbers).   The primary data store is expected to give high throughput for both sizes of workload. 

\textbf{Linear scalability}
For most cases in real world, it is difficult to provision adequate resources in advance to deal with all projected growth in content and workload. In our model, the digitized content will continue to grow as more books are added to the corpus.  This steady growth can result in insufficient storage and slow query response. Increasing workloads, especially at the 1,000,000 request, can lead to slow query response even without hosting huge amount of data. So the capability of easily scaling out through addition of additional nodes is required.

\textbf{Failure resistance}
Node failure is frequently the norm for a distributed storage system that hosts a huge amount of data across a number of nodes. When failure happens, clients should still be able to access data, preferably with no obvious perceived performance slowdown.

\textbf{Flexible schema}
Digitized volumes vary on the number of pages they contain and the fields of the structural metadata are frequently inconsistent across volumes. The document store and column-oriented store have flexible data model and match digitized volumes with their structural metadata.

\textbf{Server-side filtering}
Server-side filtering is not a strict requirement but preferred for efficiency. For example, reading all non-empty pages for a set of volumes requires filtering based on page size. In general, filtering is the efficient simple preparation of data within the data storage system to reduce the overall volume of data that is returned.

\textbf{Binding with computational framework}
Similar with filtering, binding of a storage system to a  computational framework is also a preferred feature. Binding with a computational framework like Hadoop or Spark enables moving computation to the data in cases where data sizes are large (request $>$ 300,000 volumes). The computational framework can be seen as an extra layer between the federated interface and data stores to enrich the support of queries that is more complex and analytical.  Though the computational framework is not reflected in Figure \ref{fig:htrcArch}, it can be added if the primary data store supports the binding.  

\subsection{Candidate databases}
 Cassandra, HBase and MongoDB compare differently against the requirements of HathiTrust. Specifically, we compare and contrast the three options on their scalability, failure resistance, flexible schema, server-side filtering and computational framework binding. Failure resistance is also measured experimentally, as is random access performance. 

\textbf{Cassandra}
Cassandra is a fully symmetric peer-to-peer sysyem. Data is evenly distributed across a Cassandra cluster through a hash on partition keys. The peer to peer structure allows Cassandra's linearly \textit{scalability} \cite{lakshman2010cassandra} through new nodes added new online.  Virtual nodes can balance data distribution across the cluster with heterogeneous nodes based on node's capacity.  With sufficient replication factor in addition to peer to peer structure, both data redundancy and node function redundancy can be guaranteed giving a high degree of \textit{failure resistance}.  To \textit{schema flexibility}, with latest Cassandra 3.x, dynamic columns can hold fields at page level and static column can hold metadata fields at volume level. And dynamic columns can be added and dropped at any point of time. On \textit{server-side filtering}, the Cassandra Query Language (CQL) is syntactically a subset of SQL. CQL supports $WHERE$ clause so that filtering can be pushed down to Cassandra nodes. Finally, Cassandra works with Spark in support of \textit{binding with a computational framework} by means of a connector\footnote{https://github.com/datastax/spark-cassandra-connector} that enables Spark to create node local partitions based on a Cassandra token range for data locality. 

\textbf{HBase}
HBase is an implementation of BigTable \cite{chang2008bigtable} that uses a primary/replia model.  HBase \textit{scalability} allows it to scale out by adding new nodes to host more region servers and HDFS data nodes. However, for heterogeneous nodes, data skew can occur and cause weak nodes to be overloaded and strong nodes to be relatively idle.  HBase \textit{failure resistance} depends on HDFS for data redundancy. To remove single point of failure for node function redundancy, extra standby HMaster and secondary Namenode are used. More discussion of failure resistance will be mentioned in quantitative analysis. On \textit{schema flexibility}, similar to BigTable, metadata fields can be identified with column family and column qualifier that can be added and removed at any point of time. On \textit{server-side filtering}, HBase is absent of data types \cite{gudivada2014nosql}.  Further, it does not have built-in support for simple filtering. Finally, on  \textit{binding with a computational framework}, HBase can integrate with Spark through a connector\footnote{https://github.com/hortonworks-spark/shc}; data locality can be achieved by co-locating executor and region servers.

\textbf{MongoDB}
MongoDB is a document store distributed using a primary/replica model and scales out through auto-sharding \cite{lourencco2015choosing}.  On \textit{scalability}, for a MongoDB sharded cluster, if a new node is added to a shard, the capacity of this shard scales out instead of the entire cluster. It means that the requests that fetch data from other shards does not benefit from the new node. On \textit{failure resistance}, data redundancy and node function redundancy are achieved through replica sets and multiple copies of routers and configure servers. On \textit{schema flexibility}, MongoDB as a document store uses binary serialization format (BSON) to store documents (GridFS is not considered since it does not fit the required data model and access pattern). No document structure is enforced so there is no need to declare schema before inserting data. However, the size limit of a MongoDB document is enforced to avoid excessive use of RAM by any single document.  On \textit{server-side filtering}, MongoDB supports almost all SQL operations and has built-in MapReduce for condensing large volumes of data into useful aggregated results\footnote{https://docs.mongodb.com/v3.2/core/map-reduce/}. Finally, on \textit{binding with a computational framework}, MongoDB can work with a parallel computing framework like Spark through a connector\footnote{https://github.com/mongodb/mongo-spark} through which Spark is aware of the hash range of each shard for data locality.

\begin{table*}[t]
%\resizebox{\textwidth}
\centering
\caption{Summary}
\vspace*{-4mm}
\begin{tabular}{|c|c|c|c|c|c|c|c|} \hline
&\textbf{Throughput}
&\textbf{Failure Resistance}
&\textbf{Scalability}
&\parbox[t]{1.0cm}{\textbf{Fexible Schema}}
&\parbox[t]{1.2cm}{\textbf{Simple Filtering}}
&\parbox[t]{2.55cm}{\textbf{Computational Framework Binding}} \\ \hline
\textbf{Cassandra}&\parbox[t]{2.2cm}{Second highest at both page and volume level}&\parbox[t]{2.8cm}{1) No obvious performance degradation with read repair;\newline 2) Noticeable performance drop with read repair off}&\parbox[t]{3.0cm}{Theoretically linearly scalable. Throughput does not increase linearly under added nodes with read repair on.}&Yes&\parbox[t]{1.8cm}{Moderate \newline (CQL)}&Yes\\ \hline
\textbf{HBase}&\parbox[t]{2.2cm}{Lowest at both page and volume level}&\parbox[t]{2.5cm}{System fail when 2 out of 6 nodes killed, see Figure \ref{fig:hbase-failure-resistance}} &\parbox[t]{3.0cm}{Throughput fails to increase linearly with additional nodes}&Yes&No support&Yes\\ \hline
\textbf{MongoDB}&\parbox[t]{2.2cm}{Highest at both page and volume level}&\parbox[t]{2.5cm}{Failures noticable to clients of on-going requests}&\parbox[t]{3.4cm}{Only scale out specific shards when number of new nodes to add is smaller than number of shards}&\parbox[t]{1cm}{Yes(with doc size limit)}&Strong&Yes\\ \hline
%\hline
\end{tabular}
\label{tab:summary}
\end{table*}

\section{Experimental Evaluation} \label{sec:quantitative}
Our experimental evaluation focuses on \textit{read throughput} for random access and \textit{failure resistance}. For the former, we compare throughput (as $qps$ or queries per second) of each data store at the maximum stable throughput level. For the latter, we measure change in throughput (drop of $qps$) in response to node failures.

\textbf{Read throughput.}
We test using two types of reads. Key-based record reading corresponds to random access at volume level. Access of specific fields by keys and field names, the second test, corresponds to random access for specific pages of volumes. We use the YCSB \cite{cooper2010benchmarking} benchmark tool to evaluate read throughput. 

\textbf{Failure resistance performance.}
Since failure is the norm for distributed storage systems \cite{chang2008bigtable,lakshman2010cassandra}, we evaluate resilience under node failures by manually killing nodes. We allow a data store to reach a maximum stable throughput a heavy workload, then kill a node and monitor system throughput ($qps$) through the degradation.

\textbf{Experiment data}
We use the HathiTrust corpus, a copy of which resides at the Hathitrust Research Center \cite{downie2016hathitrust}. At the time of our testing, the corpus contains 15.6 million digitized volumes. The summary of volume sizes in zipped format is shown in Table \ref{tab:tab0}. We select a sample of 180,000 items obtained by means of random sampling. 

\begin{table}[!htbp]
\centering
\caption{Volume sizes across HathiTrust collection}
\vspace*{-4mm}
\begin{tabular}{|l|l|l|l|} \hline
\textbf{Stats}&\textbf{Size (Kbytes)}\\ \hline
Min & 189\\ \hline
1st Qu & 136\\ \hline
Median &330\\ \hline
Mean &516\\ \hline
3rd Qu &664\\ \hline
Max &26,583 \\ \hline
%\hline
\end{tabular}
\label{tab:tab0}
\end{table}
\vspace*{-0mm}

\subsection{Test Environment}\label{eval}

The test environment is made up of six medium-sized VMs from the NSF funded Jetstream environment\footnote{https://portal.tacc.utexas.edu/user-guides/jetstream} used as data stores. Another 4 medium VMs are used to run YCSB 0.12.0 clients with HTRC custom workload. Each medium VM has 6 vCPUs, 16 GB RAM and 60 GB local storage. The randomly sampled 180,000 volumes are ingested into data stores. All candidate data store instances are allocated with equal memory size (4GB) and configured with a replication factor of 3 and Snappy compression. 

Using Apache Cassandra 3.9, six instances are deployed to six VMs. The $Murmur3Partitioner$ is used to uniformly distribute data across the cluster. With replication factor of three and $SimpleStrategy$, the first replica is placed based on the partitioner and the other two replicas are placed on the next nodes in clockwise direction around the ring. Compaction enhances read performance for  LSM-tree based data stores so a major compaction is run after data ingest.

The HBase installation uses HBase 1.2.4 with Hadoop 2.7.3 and Zookeeper 3.4.9. Again, six medium VMs are used to host all HBase components. Each VM runs an HBase region server and an HDFS data node. Zookeeper is deployed on 3 out of the 6 medium VMs. An HMaster and a primary namenode collocate on same VM. A backup master and a seconday namenode are hosted on another VM. Replication factor 3 is guaranteed through HDFS. A major compaction is run after all data is ingested to optimize read performance. 

MongoDB version 3.4 is used with the WiredTiger storage engine. Because of MongoDB's sharded replica set model, two shards are deployed with hashed (MD5) sharding for even data distribution across the sharded cluster. Three VMs are used for each shard as replica sets to guarantee three replicas. Three VMs are chosen to host Config Servers. All six VMs runs mongos/router daemons to accept requests. The $id$ field of MongoDB is indexed with BTree.

\begin{table}
\centering
\caption{Maximum Stable Throughput for Reads}
\vspace*{-4mm}
\begin{tabular}{|c|c|c|} \hline
Data Store&Volume Level Read $qps$&Page Level Read $qps$ \\ \hline
Cassandra & 1,665 & 8,560\\ \hline
HBase & 1,040 & 7,402\\ \hline
MongoDB & 2,166 & 9,373\\ \hline
%\hline
\end{tabular}
\label{tab:max-status}
\end{table}

\subsection{Volume Level Random Access}
We first evaluate random access of volumes, in other words, whole book access, for each of the three data storage systems. We deploy four YCSB clients. The number of threads are adjusted to tune read workloads. Each YCSB client loads the list of keys (volume identifiers), shuffles the list and sends key-based requests synchronously one after another without any throughput limit. Each request is returned the record for its specified key.

Cassandra clients are configured with a token aware load balancing policy to send requests to the node that has a local copy of requested data to minimize internal traffic overhead. The read consistency level is configured to 1.   HBase clients are region-server aware. By default, they contact the Zookeeper ensemble to get the $META$ table location and cache $META$ table that has region and region server information. Then the subsequent requests contact the region server serving that region directly without going through lookup process.   MongoDB clients by default use a read preference model $primary$, that is, all read requests only hit the primary replica set of each shard. This does not fully utilize all nodes so we configure it to $nearest$ to evenly distribute requests to all nodes for each shard.

\begin{figure}
\centering
\includegraphics[scale=0.66]{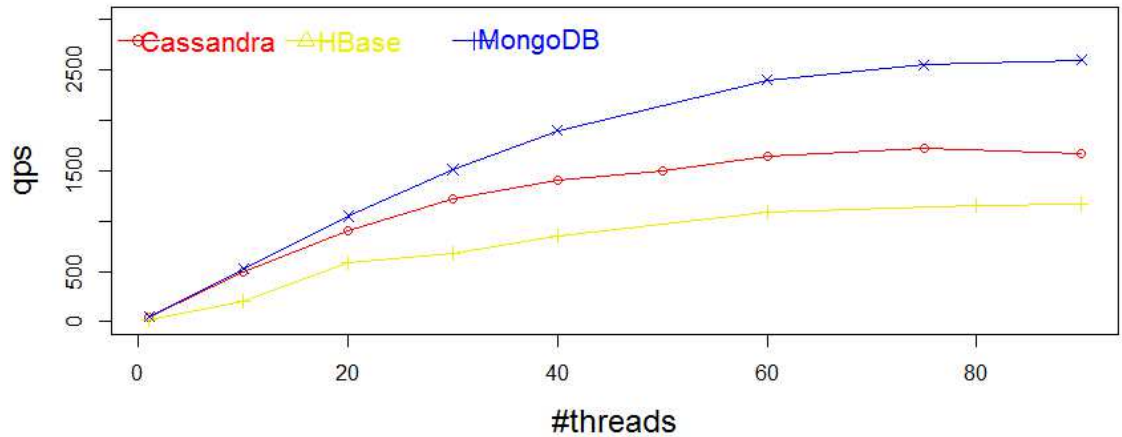} 
%\vspace*{-0.7cm}
\caption{Record/volume level read throughput.  As a function of a growing workload, record read throughput increases to a maximum stable level for each system.MongoDB outperforms HBase by 108.2\%.}

\label{fig:volume-throughput}
\end{figure}

Read throughput at the record/volume level in queries per second (qps), see Figure \ref{fig:volume-throughput}, shows MongoDB outperforming Cassandra and HBase by 30\% and 108.2\% respectively. Cassandra outperforms Cassandra by 37.5\%. And the maximum stable throughput for each is listed in column two of Table \ref{tab:max-status}.

\subsection{Page Level Random Access}
\begin{figure}
\centering
\includegraphics[scale=0.7]{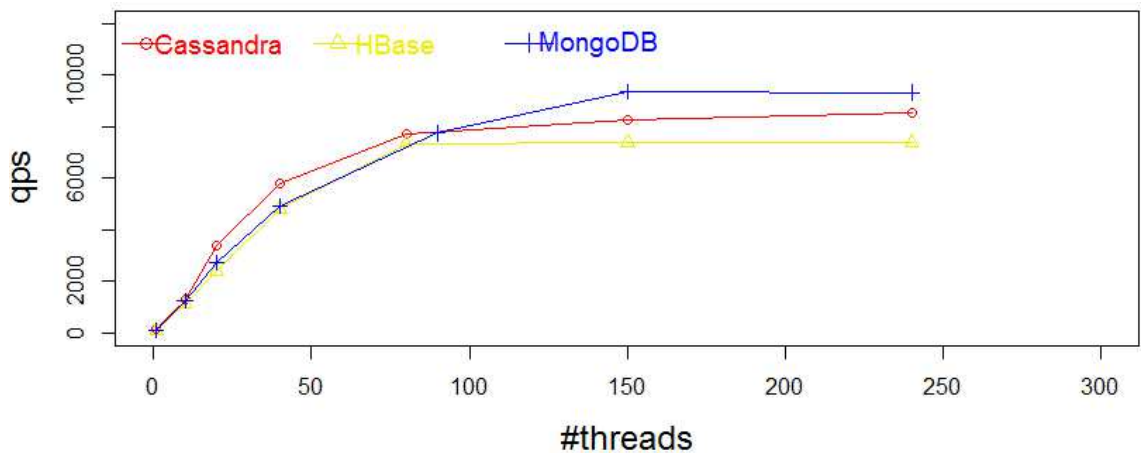} 
%\vspace*{-0.7cm}
\caption{Page (field) level read throughput. As the workload increases, field read throughputs of data stores increase accordingly and attains a maximum stable status at 150 threads;
MongoDB outperforms HBase by 26.6\%}
\label{fig:page-throughput}
\end{figure}

We evaluate read performance of  page-level content. Each YCSB client loads a list of keys, shuffles the list, and sends key-based requests with 10 specified fields (page sequences) to fetch. The 10 field names, in this case 10 page sequences, are generated randomly between 1 and 300. Requests are sent one after another with uniform distribution. The experiment result for field read in Figure \ref{fig:page-throughput}, shows that MongoDB performs best and Cassandra performs better than HBase at the record/page level. The page/record level maximum stable throughputs are shown in column three of Table \ref{tab:max-status}. 

\subsection{Failure Resistance} \label{failure-resistance}
We test failure resistance, that is, a measure of a system's resiliency in the face of node level failures.  We use a volume level workload and push each NoSQL data store to its maximum stable status. After 100 seconds at maximum stable status, we manually kill a node and measure the performance in terms of overall throughput. We then allow the system to run for another 100 seconds before killing a second node. Time interval is chosen based on our observation that performance can stablize in 100 seconds for our experiments. Because the replication factor is 3, client requests are expected to be responded to with at least one available replica. 

%\begin{figure}
%\centering
%\includegraphics[width=\linewidth]{figures/cassandra-failure-resistance.jpg} 
%\vspace*{-0.7cm}
%\caption{Failure resistance test, Cassandra. The first node is killed at 100th second and the other at 200th second. The result does not show obvious performance degradation when at least one replica available.}
%\label{fig:cassandra-failure-resistance}
%\end{figure}

\begin{figure}
\centering
\includegraphics[scale=0.75]{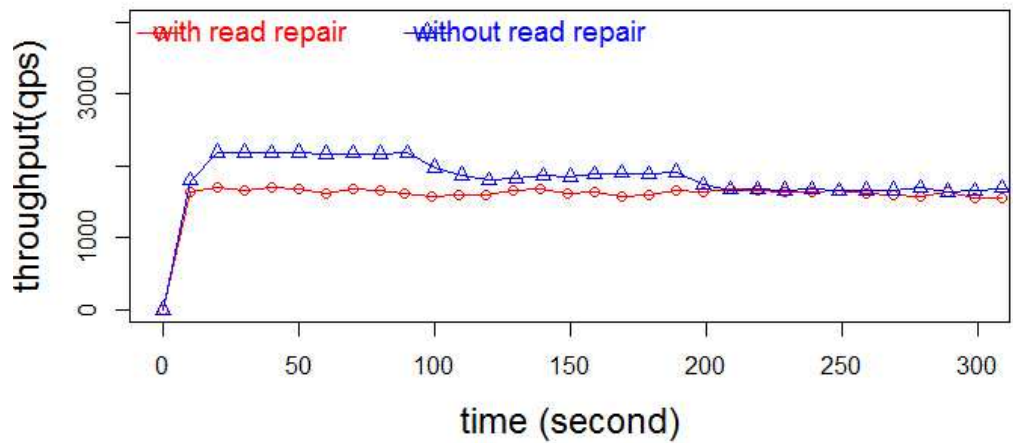} 
%\vspace*{-0.7cm}
\caption{Impact of read repair; Cassandra. When read repair is disabled, the maximum stable throughput is higher and degrades with node failures. With read repair, the maximum stable throughput is lower and does not have obvious degradation with node failures.}
\label{fig:read-repair}
\end{figure}

\textit{Cassandra}
Because of the peer-to-peer architecture that is fully symmetric, there is no difference in which nodes to kill. So two random nodes are picked. The result of Cassandra failure test is shown in red (with read repair) in Figure \ref{fig:read-repair}. We can see no obvious performance degradation. 

The reason why there is no obvious performance degradation is related to replication factors and anti-entropy process of Cassandra, a process of comparing the data of all replicas and updating each replica to the newest version. Even for read consistency level $ONE$, a portion of read requests are chosen for read repair, which causes the overhead for overall read throughput. When a node is killed, the lost node makes throughput lower but on the other hand, the reduced number of replicas increase the throughput, therefore no obvious throughput changes happen. If we turn off read repair, then we can see the overall throughput are higher at saturation level and when nodes are killed, there are obvious performance degradation as shown in Figure \ref{fig:read-repair}. Since anti-entropy process is an important component for Cassandra consistency model, its performance impaction should be taken into consideration. 

\begin{figure}
\centering
\includegraphics[scale=0.7]{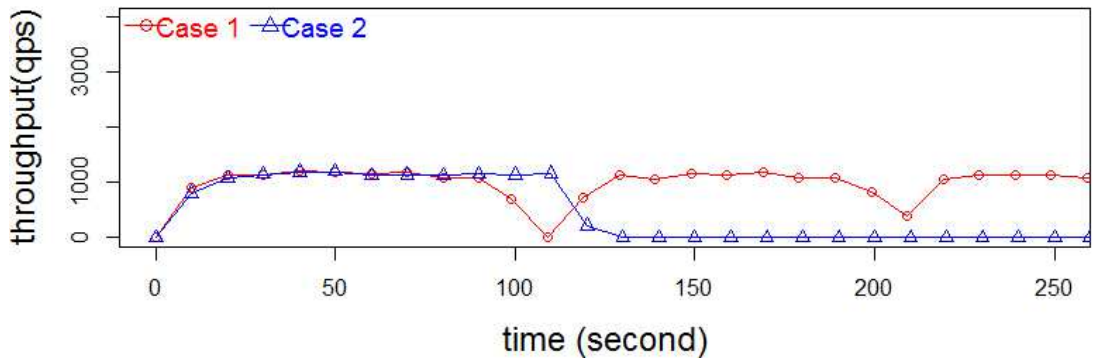} 
%\vspace*{-0.7cm}
\caption{For HBase, 1) When the first data node and region server are killed at 100 (seconds), throughput drops significantly and then rises to previous level; when the second data node and region server are killed at time 200, there is obvious performance degradation and then a rise to a lower level compared to initial maximum level by 15\%. 2) When the active HMaster \& primary namenode are killed at 50 seconds, there is no performance impact; when a region server is killed at 100, the entire cluster fails to serve any request.}
\label{fig:hbase-failure-resistance}
\end{figure}

\textit{HBase}
As HBase consists of many different components, we test two combinations of components to kill. \textit{Case 1}: kill 2 random nodes that have only region servers and data nodes running. \textit{Case 2}: kill only the active HMaster and primary name node processes at 50 seconds without touching any data node and region servers; then kill a region server at 100 seconds. Figure \ref{fig:hbase-failure-resistance} shows the results. 
We can tell that when one of six data nodes and region servers are killed, the overall performance drops and then rises. When the second data node and region server are killed, the performance also first drops to some point and rises back up but to a lower level compared to maximum stable level by 15\%.  

The result of the second HBase failure test is more complex. When active HMaster and primary namenode are killed, backup HMaster is notified to become active and secondary namenode is notified to be primary; backup HMaster tries to connect to HDFS while the secondary namenode has not become primary yet so backup HMaster throws uncatched exception and crashes; after a short while, secondary namenode successfully become primary. However, this entire process does not affect client side because client side has $META$ table cached at the first request. Then after a region server is killed, Zookeeper sees the failure through heartbeats and requires clients to halt until a updated $META$ table is available. HMaster should also be notified to reassign regions served by killed region server to other active region servers and update $META$ table. Since both active and backup HMasters are down, region reassignment and $META$ update never happens. Therefore, the entire system is not accessible and gives 0 throughput.  

\begin{figure}
\centering
\includegraphics[scale=0.7]{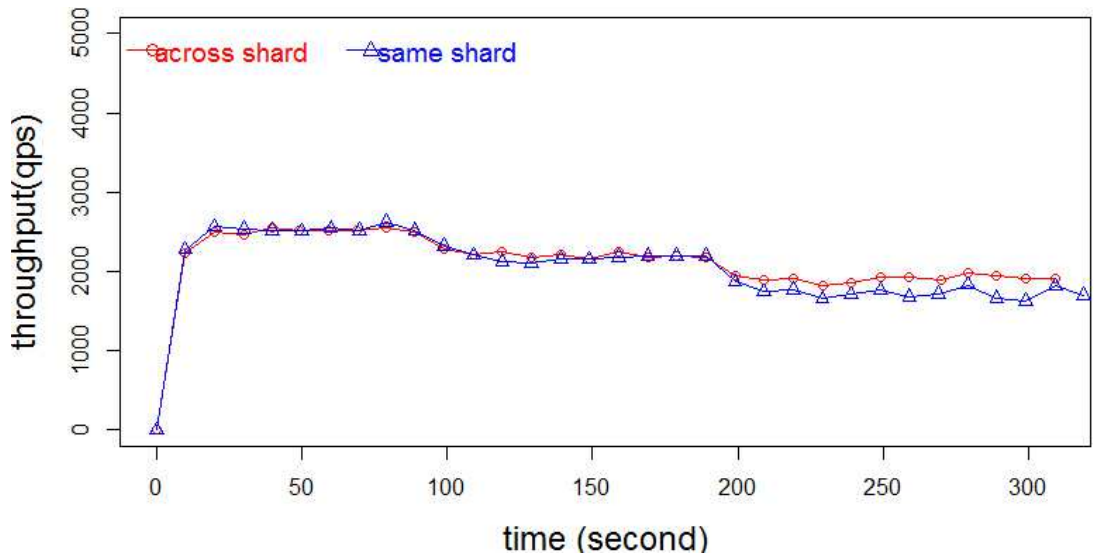} 
%\vspace*{-0.7cm}
\caption{For MongoDB, irrespective of whether the killed nodes are from the same shard or different shards, overall throughput drops by 1/6 under node failure.}
\label{fig:mongo-failure-resistance}
\end{figure}

\textit{MongoDB}
For MongoDB, also two different cases are tested. The first case is to randomly choose two nodes from the same shard to kill. The other case is to choose one random node from each shard to kill. The results are shown in Figure \ref{fig:mongo-failure-resistance}. The results shows that there is a predictable pattern of performance degradation for MongoDB, that is, overall performance in terms of throughput drops by 1/6 of the original maximum stable level. Though the performance degradation is more obvious compared to Cassandra and HBase, MongoDB still has better read throughput even when 2 nodes out of 6 are down. However, from perspective of failure masking of MongoDB, clients do get query exceptions for ongoing requests at the moment when nodes are killed, which would need a higher layer of service to help.

\section{Discussion} \label{discussion}
All three systems discussed have a flexible schema and tight binding with a large scale computation framework. However, from the perspective of read throughput for random access, failure resistance, scalability and server-side filtering, the data stores behave quite differently:

MongoDB has the best read performance at both record/volume level and field/page level. The sharded architecture requires more nodes to scale out the entire cluster. When node failure occurs and each shard maintains working replicas, read performance drops and clients notice exceptions. To hide exceptions, a higher layer of service above MongoDB is needed. MongoDB has the best support for server-side filtering. 

Cassandra has the second best read throughput. It makes use of any number of extra nodes to scale out the entire cluster. The anti-entropy mechanism can cause internal traffic and thus slow down the overall read throughput but, as a side effect, makes the node failure unnoticeable when there is at least one replica online. Cassandra has limited capability for server-side filtering through CQL. 

HBase has the lowest read throughput. Any number of extra nodes can be added as datanode/regionserver to scale out the cluster. However, based on our experiments, five datanode/region-servers gives almost identical maximum throughput to six servers even when the regions are assigned evenly. For failure resistance, when datanode/region-servers are down, clients experience a short-term performance drop. A rare case that some region server goes down after HMaster and primary namenode crash at the same time could make the entire cluster not accessible. HBase has no support for server-side filtering so a higher layer of service is needed for the filtering requirement. 

MongoDB gives the highest random access throughput at both volume and page level but its distribution architecture requires more nodes to scale out compared to Cassandra. MongoDB has a document size limit (16 MB for version 3.x) to avoid excessive use of RAM by any single document. When adding more nodes is not a problem and document sizes are smaller than 16 MB, MongoDB is the best choice. This size limit is expected to increase but, for now, we do have digitized volumes over 16 MB as shown in Table \ref{tab:tab0}. 

With a modest compromise on read performance, Cassandra is the overall choice for the HathiTrust Research Center on the basis of its ease in scalability and its superiority in masking failures all the while still giving good read performance and support for server-side filtering. See again Table \ref{tab:summary} for a summary of features of the three data stores.

\section{Use case} \label{sec:usecase}
We describe a practical use case of how researchers engage in custom analysis of the HathiTrust collection in non-consumptive way. We also show how the previous studies can be simplified with the proposed architectural model. 

Murdock \textit{et al.} \cite{murdock2015topic} uses an HTRC Data Capsule for Latent Dirichlet Allocation (LDA) topic modeling of thousands of texts.  The Capsule position in the architecture model of Figure \ref{fig:htrcArch} is at the top as "review-guaranteed non-consumptive".   The overall workflow carried out in the Capsule is to 1) gather a list of volume identifiers by searching an Apache Solr full text index, then using a tool that assembles the texts into a \emph{workset},  2) using the workset, for each volume identifier, pull its text into the Capsule (which is now blocked for any kind of network access), 3) run LDA topic model against the retrieved volumes.  4) results are available for visualizing within the Capsule or exporting for visualization outside the Capsule.  Under the proposed architectural model, steps 1) and 2) can be combined into one step.   Further, rights information is also available which are critical for science reproducibility.  

Under the hood, for a query $q$ that retrieves volume contents with metadata-based search criteria, if the number of volumes in query result is 10,000, the bibliographic metadata store will be queried only once to return 10,000 identifiers; the Rights database will be queries 10,000 times to get rights information for final result review required by non-consumptive restriction; and the primary data store is responsible for efficiently and reliably transferring the textual contents of these 10,000 volumes to the secure environment (Data Capsule). For 10,000 key-value lookup against rights database and streaming of textual content for 10,000 volumes, optimization like request batching can be used following best practice for different data stores to improve overall performance performance. Researchers need not have to know multiple back end data models and learn different sets of APIs for different data stores to follow the best practice since the entire back end is transparent and the ordering of executions of sub-queries can be optimized for researchers. 

\section{Conclusion \& Future Work}\label{sec:future}
Big data mixed-restricted collections pose unique challenges on data management.  We propose an architectural model for multi-billion page, mixed-restricted digitized content that protects the data while allowing researchers to bring their analytical computations to the data.   We undertook the practical exercise of evaluating storage systems that the proposed architecture depends on to provide reliable and efficient access. 

There are several directions of future work: firstly, further study of different data models for the federated (polyglot) interface is possible as more than one NoSQL data model that aggregates over federated resources is possible. Too, because of the size of the corpus and restrictions on its digital content, researchers move their analysis to a virtual secure container.  Performance could be improved if the secure environment can more closely co-locate with the storage system to better utilize data locality. Finally, the impact of bifurcated workload was generalized into random read but was not given extensive study here since the dominant resource for 1,000 volume read and 300,000 volume read can be different and result in different performance. So bifurcated workload impact can be further studied at scale rather than the six medium JetStream VM cluster used in this study.

%\section{Acknowledgements}
%Thanks to Yu (Marie) Ma and Milinda Pathirage for participation in the data storage architecture. The outcome of this study was eventually used in the HathiTrust Research Center, thanks to their engaging discussions.  This work funded in part by the HathiTrust Consortium and the Andrew W. Mellon Foundation under grant #41500672.  

%\nocite{*}% Show all bib entries - both cited and uncited; comment this line to view only cited bib entries;
\bibliography{wileyNJD-AMA}%

%\section*{Author Biography}

%\begin{biography}{\includegraphics[width=66pt,height=86pt,draft]{zong.pdf}}{\textbf{Zong Peng.} Zong Peng received the Ph.D. degree in Computer Science from Indiana University Bloomington in 2018. He worked in Data To Insight Center as research assistant and was advised by Prof. Beth Plale. This study is undertaken as part of his dissertation. His main areas of research interest are full text search engine, big data management and text mining. He is currently an engineer of Microsoft AI and Research.}
%\end{biography}

%\begin{biography}{\includegraphics[width=66pt,height=86pt,draft]{plale.pdf}}{\textbf{Beth Plale.} Beth Plale is founding Director of the Data To Insight Center and founding Director of HathiTrust Research Center.  She is a full professor in the School of Informatics, Computing, and Engineering at Indiana University Bloomington.  Professor Plale's research interest are in Big Data, cloud computing, data sharing, persistent IDs for data reuse, and Open Science.}
%\end{biography}

\appendix
The bulk of HTRC data is digitized textual content. For each volume, the original raw format is a zip file named by its volume id ($<volumeId.zip>$). Each zip entry is a page of the volume named by its sequence number($pageSequence.txt$). 

In experiments listed in paper for choosing storage system for HTRC, Cassandra, HBase and MongDB are three candidates. A sample of HTRC textual data is ingested into these storage systems as test data set for efficiency test at both volume level and page level random access.
\paragraph{Cassandra Data Model Details}
For Cassandra, HTRC data is stored in Cassandra column family in the following data model for both volume level random access and page level random access. 

\begin{enumerate}
    \item Volume ID is used as partition key. 
    \item page sequence is used as clustering key.
    \item volume level metadata (e.g. volume size) is stored in static columns. 
    \item page textual contents are stored in dynamic column "contents".
\end{enumerate}

To be more specific, the CQL statement to create a Cassandra Columnfamily used by experiments in paper can be written as below:

\begin{lstlisting}
String createTableStr="CREATE TABLE " + columnFamilyName + " ("				    	       + "volumeID text, "
    + "accessLevel int static, "
    + "language text static, "
    + "volumeByteCount bigint static, "
    + "volumeCharacterCount int static, "
    + "sequence text, "
    + "byteCount bigint, "
    + "characterCount int, "
    + "contents text, "
    + "checksum text, "
    + "checksumType text, "
    + "pageNumberLabel text, "
    + "PRIMARY KEY (volumeID, sequence))";
\end{lstlisting}

For volume level random access, reads are based on partition key. For page level random access, reads are based on both partition key and clustering key. 
\paragraph{HBase Data Model Details}
For HBase, HTRC data is stored in HBase table in the following data model for both volume level random access and page level random access.
\begin{enumerate}
    \item Volume ID is used as row key. 
    \item Page contents are organized under column family "Contents". 
    \item Page sequences are used as column qualifiers. 
    \item page textual contents are stored in cells located by volume id, "contents" column family and "sequence" qualifier.
\end{enumerate}

For volume level random access, reads are based on row key and column family name. For page level random access, reads are based on both row key, column family name and specified qualifiers.
\paragraph{MongoDB Data Model Details}
For MongoDB, HTRC data is stored in MongoDB collection with an embedded document model for both volume level random access and page level random access. Each MongoDB BSon document is a HTRC volume with the following data model. 
\begin{enumerate}
    \item Volume ID is used as custom MongDB \_id. 
    \item Each page is also a document embeded in the volume document with page sequence as field name. 
    \item Within page document, page textual contents are stored in "contents" field along with other page level metadata (e.g. checksum).  
\end{enumerate}

For volume level random access, reads are based on document \_id. For page level random access, reads are based on both document \_id and page sequence.

\end{document}